# Des activités de recherche au pilotage de l'IRD : les enjeux de la modernisation d'interfaces et de l'interopérabilité des données

S. Tostain

## 1 Résumé

Les activités de la recherche sont généralement observées et évaluées à travers le prisme de leur production et des éléments financiers ou de compositions des équipes. En plus des indicateurs de gestion standardisés et bibliométriques, l'IRD poursuit depuis une dizaine d'années la construction de nouveaux indicateurs, sur la base des déclarations annuelles réglementaires des chercheurs de l'Institut. Différents outils du management qualité permettent l'évolution des différentes interfaces. Cette source de données, plus « ouverte » et plus « utile » par son intégration dans le système d'informations de l'Institut est adaptée aux besoins du pilotage pluriannuel de la recherche à l'IRD. La finalité est double : (1) progresser dans l'évaluation de la recherche et dans la maîtrise de l'information par tous les acteurs, (2) éclairer précisément le plus d'acteurs possibles via des circuits et des outils numériques plus efficients. L'objet du présent article est d'exposer la conduite à l'IRD du changement sur toute la chaîne de production et sur les indicateurs d'activités des chercheurs et chercheuses pour une meilleure cartographie des activités scientifiques.

## 2 Introduction

Des canaux d'irrigation existent-ils vraiment et sont-ils la preuve d'une vie intelligente sur Mars ? A la fin du XIXe siècle, de nombreux scientifiques ont répondu « oui » sur la base d'observations au télescope et d'une cartographie interprétée de la planète Mars [1]. A l'époque, ces images suggéraient des alignements droits et croisés, similaires à un système d'irrigation, avec une interprétation possible : il y a des canaux, de l'eau et donc des traces de vie intelligente sur Mars. On sait depuis, grâce aux images de haute résolution et des missions exploratoires sur le sol martien, qu'aucun système de canaux artificiels n'est présent sur cette planète. Les scientifiques sont devenus moins myopes avec des outils de plus « haute résolution ». Les observations directes et indirectes ont fourni depuis des données plus riches et surtout plus précises pour interpréter.

Dans le cadre du pilotage institutionnel des établissements de recherche, l'observation fine des moyens, des activités et de la production des équipes de recherche a pour enjeu d'établir ou d'ajuster la vision réelle, d'évaluer les impacts de différentes actions vis-à-vis des objectifs fixés ou à prévoir [2]. Comment veiller à l'atteinte d'un objectif de développement et de formation de nouvelles générations de scientifiques dans les pays du Sud si aucune information n'est demandée aux acteurs de la recherche sur la nationalité des étudiants, les établissements d'inscription des doctorants accompagnés sur plusieurs années ? La fiabilité et la richesse des données-sources, leur adaptation au spectre le plus large possible de questions posées, sont à la racine de tout flux pérenne d'organisation de « la donnée » réutilisable et « emboîtable » sur le modèle des poupées russes. L'analyse multi-résolution des données - de la maille la plus fine à la plus macroscopique – est un exercice d'équilibriste constant pour l'institution : quels niveaux de détail demander aux chercheurs pour répondre de manière synchrone à des questions multiples posées par les instances ou les directions institutionnelles (critères d'observation et niveaux de précision variables attendus des données ou des résultats de calculs) ?

La mise en œuvre de ce flux d'informations et la mise à disposition des données ou des indicateurs nécessite une connaissance approfondie des métiers de la recherche d'une part et de ceux du pilotage de la recherche d'autre part. L'évolution des questions, la maîtrise du traitement des données et la restitution sont à la source des travaux d'adaptation et d'amélioration continue menés ces dix dernières années à l'Institut de recherche pour le développement (IRD). L'objectif global est de disposer d'une vision fine et transversale des activités quotidiennes des chercheurs et chercheuses en accord avec les besoins de l'évaluation individuelle ou collective et des besoins de la gouvernance : deux types d'usagers de l'information (ou « clients ») qui nécessitent une adaptation continue des supports, la rationalisation des flux d'informations centralisées, la modernisation des interfaces et des outils (de collecte, d'accessibilité ou de la mise en valeur des informations sous la forme d'indicateurs de pilotage et l'identification d'experts de l'IRD).

L'objet de cet article est de partager l'expérience globale et continue de l'IRD sur les activités décrites chaque année par ses chercheurs et ses chercheuses, qui permettent d'éclairer le pilotage scientifique et institutionnel. *L'enjeu est de diminuer la myopie instrumentale et d'étendre le spectre des observations pour s'adapter* aux nouvelles questions, permettre des interprétations éclairées, reformuler des priorités et identifier les moyens de faire levier si nécessaire pour atteindre les objectifs fixés à l'institut.

Après une présentation du contexte institutionnel de l'IRD et de l'existant, le périmètre global du projet sera plus spécifiquement identifié. Différents processus imbriqués sont en construction continue avec plusieurs phases d'évolutions majeures. Quelques clés et outils construits pour le suivi du projet et la programmation des actions sont présentés dans un dernier chapitre pour illustrer concrètement ce partage d'expérience. En conclusion, les points forts du projet seront soulignés et les activités de modernisation mises en perspectives par rapport aux enjeux.

## 3  Le contexte

L'IRD est un établissement public de recherche pluridisciplinaire ((Établissement public à caractère scientifique et technologique, EPST), sous double tutelle ministérielle de l'enseignement supérieur et de la recherche (MESR) et de l'Europe et des affaires étrangères (MEAE). Ses domaines de recherche, toutes disciplines scientifiques confondues, sont liés aux objectifs de développement et de Science durable dans les pays du Sud, portés par des enjeux scientifiques internationaux [3]. Ses scientifiques travaillent en France comme à l'international dans des contextes institutionnels variables, ce qui entraîne une forte complexité à observer finement les activités d'équipes dispersées et à quantifier ses propres innovations et impacts (directs ou indirects) [4]. L'adaptation continue aux paysages institutionnels et géopolitiques, avec la mise en œuvre de moyens adaptés pour servir de levier à l'atteinte de ses objectifs, a renforcé le besoin de production plus fluide et plus rapide d'indicateurs pérennes sur les activités de ses équipes, en plus des indicateurs standards de gestion administrative.

En observant les indicateurs affichés par l'établissement, on constate des choix différents des chiffres-clés fournis selon leur niveau de visibilité. Du site internet au rapport annuel d'activités jusqu'au contrat d'objectifs et de performances de l'Institut, tous les indicateurs suggèrent des sources de données identiques, organisées en « système gigogne » pour des calculs à des échelles différentes, dans le temps. Par exemple, le montant global du budget annuel de l'IRD (240 M€ de budget en 2019, 26,6 M€ de ressources sur contrats de recherche), est construit sur la même source de données financières du système d'information que celui du montant des bourses accordées à des étudiants du Sud par l'institut. Ce dernier est en revanche présenté sous l'angle du nombre de bénéficiaires de ces bourses (128 doctorants avec allocation de recherche pour une thèse au Sud en 2019) plutôt que le montant global alloué : deux axes et niveaux de détails sont ainsi possibles pour illustrer les activités à partir des mêmes données-sources. Ces chiffres-clés sont par définition des indicateurs de « gestion », de par leur nature et de par l'origine commune de leurs calculs.

Des indicateurs supplémentaires sont recherchés pour mesurer en continu, évaluer, piloter ou valoriser les impacts des activités, à l'IRD comme ailleurs [2]. En parallèle, se développent dans les entreprises privées des indicateurs de capital humain (ICH) pour mesurer l'efficacité ou les performances des ressources humaines en tant que capital interne et ils sont utilisés comme aides à la décision [5]. Dans les EPST, les indicateurs de « performances » des scientifiques ont pour leur part été longtemps illustrés essentiellement par des indicateurs bibliométriques, résultats de calculs sur leur « production » [6] (exemples affichés de l'IRD dans le rapport d'activités 2019 : 5621 publications sur le périmètre des unités mixtes de recherche dont au moins un signataire est de l'IRD en 2019, 1 687 références d'articles publiés en 2021 par les chercheurs de l'IRD dans le *Web of Science*, 63% de co-publications avec un partenaire du Sud). Les autres produits et innovations de la recherche sont les brevets et les instruments développés et diffusés. Les universités présentent d'autres indicateurs « naturels » de suivi et valorisation de leurs activités : le nombre d'étudiants et le nombre de diplômés, preuves de résultat performant de leurs missions d'enseignement supérieur et de transmission des savoirs académiques [7].

Selon la ou les finalités d'usage des indicateurs à produire – et donc à construire - doivent donc être bien identifiés : a) pour qui et pour quoi faire, b) le détail des informations initiales nécessaires et leurs sources avec la qualification des « objets » d'observation, c) l'identification de la maille minimale et la périodicité les plus pertinents pour les produire. Les données-sources pour la production d'indicateurs de pilotage de nature pérenne sont stockées : elles « restent » mais les analyses, les critères et les questions posées changent dans le temps pour s'adapter aux besoins. Le « comment produire » inclue le besoin d'imbrication des données pour des indicateurs multi-échelles, avec la nécessité d'*ouverture* et d'*interopérabilité* des données et des outils.

Comme pour les données et les résultats de la recherche *stricto sensu* auxquels s'appliquent les concepts de *Science ouverte* : cahiers de laboratoires numériques, maîtrise des méthodes et traçabilité des résultats d'expériences ou d'enquêtes, des échantillons, des « données » collectées en amont de publications scientifiques, mutualisation de données ou de moyens de calculs, etc. La transformation et la modernisation numérique dans la maîtrise des données d'activité individuelle des chercheurs pour le pilotage scientifique est en cours à l'IRD. Elles sont réalisées dans un esprit de « partage » utile au plus grand nombre dans une base de données fiables et descriptives sur les activités des chercheurs répondant aux principes numériques *FAIR* (*F*indability, *A*ccessibility, *I*nteroperability, and *R*euse[1]) [8].

Le schéma directeur du numérique de l'IRD met l'accent sur la modernisation du système d'informations et une architecture interopérable. La production d'indicateurs de pilotage de l'activité scientifique de l'IRD entre dans ce cadre. Il complète le panel des indicateurs de pilotage existants et des indicateurs bibliométriques maîtrisés par ailleurs. Les activités multiples des chercheurs et des chercheuses de l'Institut sont décrites au plus réel et au plus précis dans leurs supports individuels et réglementaires [9;10]. Parmi ces supports, seule la *fiche annuelle d'activités (FAA)*[2] est construite depuis une dizaine d'années à l'IRD sur un format de l'information exploitable en tant que donnée-source structurée. L'ensemble des fiches annuelles individuelles constituent la source déclarative à la base de données interne et institutionnelle du même nom « FAA ».

Le *projet FAA* est en trois volets :
- les questions posées aux chercheurs et les réponses, données-sources pour décrire l'activité scientifique ;
- leur stockage et leur accessibilité pour davantage de transversalité dans leurs usages ;
- la production finale d'indicateurs annuels et multicritères, permettant aussi la valorisation des expertises spécifiques l'IRD.

## 4  Le projet FAA et son système

L'acronyme « FAA » pour la Fiche Annuelle d'Activités résume d'une part le processus annuel obligatoire pour tou(te)s les chercheur(euse)s titulaires [10], chargé(e)s ou directeur(trice)s de recherche (environ 800 chaque année). Il identifie d'autre part le support lui-même pour réaliser l'exercice à l'IRD et décrire l'essentiel de leurs missions statutaires : « développement des connaissances, leur transfert et leur application dans tous les domaines contribuant au progrès de la société, la diffusion de l'information et de la culture scientifique et technique dans toute la population, et notamment parmi les jeunes, la participation à la formation initiale et à la formation continue, l'administration de la recherche » [9].

---

[1] Identifiables et retrouvables, accessibles, interopérables et réutilisables ou reproductibles
[2] Equivalent au CNRS, en-dehors de son contenu ou de son format, du CRAC (compte rendu annuel d'activité des chercheurs) ou du RIBAC (Recueil d'Informations pour un oBservatoire des Activités de reCherche en SHS).

La base de données « FAA » est conservée et exploitée jusqu'à présent à la MEPR[3] pour répondre à ses missions d'aide à la décision de la Gouvernance de l'IRD et de l'évaluation de la recherche. Le « projet FAA » englobe le suivi d'activités et l'évolution continue de trois volets aux enjeux interdépendants :

- le *recueil des informations* sur les activités des chercheurs(euse)s et les réponses aux questions posées. Il y a aussi l'aide aux utilisateurs pour simplifier la déclaration réglementaire et faire évoluer les questions pour répondre aux besoins institutionnels d'observation des activités ;
- le *traitement et cohérence des données* mobilisant d'autres services institutionnels selon les sujets et leurs équipes à former et à encadrer pour maîtriser les données et les flux, prévoir leur interopérabilité, respecter la réglementation (décret n°83-1260, RGPD, confidentialité) ;
- la *restitution et valorisation des données* sous forme d'indicateurs et analyses croisées pour conseiller ou pour orienter les autres services institutionnels sur leurs propres supports et le pilotage de la donnée, construire des indicateurs plus « robustes », formuler des questions, identifier et élargir les usages de données transversales. Le tout facilite l'interrogation et enrichit l'offre de données chiffrées pluriannuelles de l'IRD et accélère la réponse à des demandes d'indicateurs, d'éléments de feuille de route ou de stratégies.

Dans le langage courant des qualiticiens, deux groupes de « clients » - au sens de la norme ISO 9001 [11] principaux sont identifiés pour ce système dans le périmètre du projet : l'un en début de chaîne de l'information (*les chercheurs et chercheuses* qui fournissent de la donnée déclarée), le second qui exploite le contenu de la base de données d'informations par de la consultation (les décideurs, évaluateurs ou « *pilotes* »). L'innovation méthodologique principale dans la gestion du projet FAA est de considérer le chercheur comme « client » de ses propres données et comme « client » pilote-décideur au sein de l'établissement : il doit bénéficier et percevoir aussi positivement que les pilotes de l'institution tout changement et amélioration dans l'accès et la maîtrise des données. Ses déclarations annuelles successives font parties des pièces du grand puzzle constitué par la base de données centralisée du même nom « FAA » (voir Fig. n°1).

## 5   Les évolutions du système pour produire des indicateurs d'activités des chercheurs IRD

Comme souligné en introduction et dans les éléments de contexte, l'objectif est de pouvoir décrire et informer plus précisément les acteurs du pilotage de l'institut tout en permettant à tous les chercheurs

---

[3] MEPR : Mission évaluation et programmation de la recherche, rattachée à la Direction du pôle Science de l'IRD

de répondre à l'obligation réglementaire annuelle de déclarations de leurs activités. Au cœur de cet objectif : des données bien organisées décrivant le plus finement possible les activités décrites pour répondre aux questionnements, soit par leur niveau de détails suffisant, soit par croisement avec d'autres critères et informations connues de l'établissement présentes par ailleurs dans le système d'informations. Durant ces dix dernières années, différents changements ont été repérés et/ou opérés à l'IRD sur tout ou partie du système, des outils et des interfaces associées pour répondre à cet objectif et l'intégration progressive de FAA au système d'information avec, entre autres, la mise en œuvre de référentiels transverses à plusieurs applications internes.

Les cahiers de l'école qualité et responsabilité sociétale en recherche et en enseignement supérieur – Septembre 2021 – *in press 2023*

Figure n°1 : Schéma global du système couvert par le projet « FAA » avec deux interfaces-clients identifiées selon leurs rôles et leurs besoins vis-à-vis de l'information centralisée des contenus de fiches annuelles d'activités des chercheurs/euses de l'IRD. L'interface avec les chercheurs à gauche est en cours de transformation avec passage d'un format Excel à un formulaire en ligne, l'interface à droite avec les pilotes institutionnels et la communauté IRD sera améliorée par la suite. Des copilotes de la donnée sont déjà mobilisés dans la maîtrise de l'interface avec les chercheurs et la base de données, et une interface de consultation d'indicateurs pluriannuels sur une partie des données en base a été construite antérieurement (« Azimut »). Ces changements sont décrits plus précisément dans la suite du présent article (source : IRD Sabine Tostain).

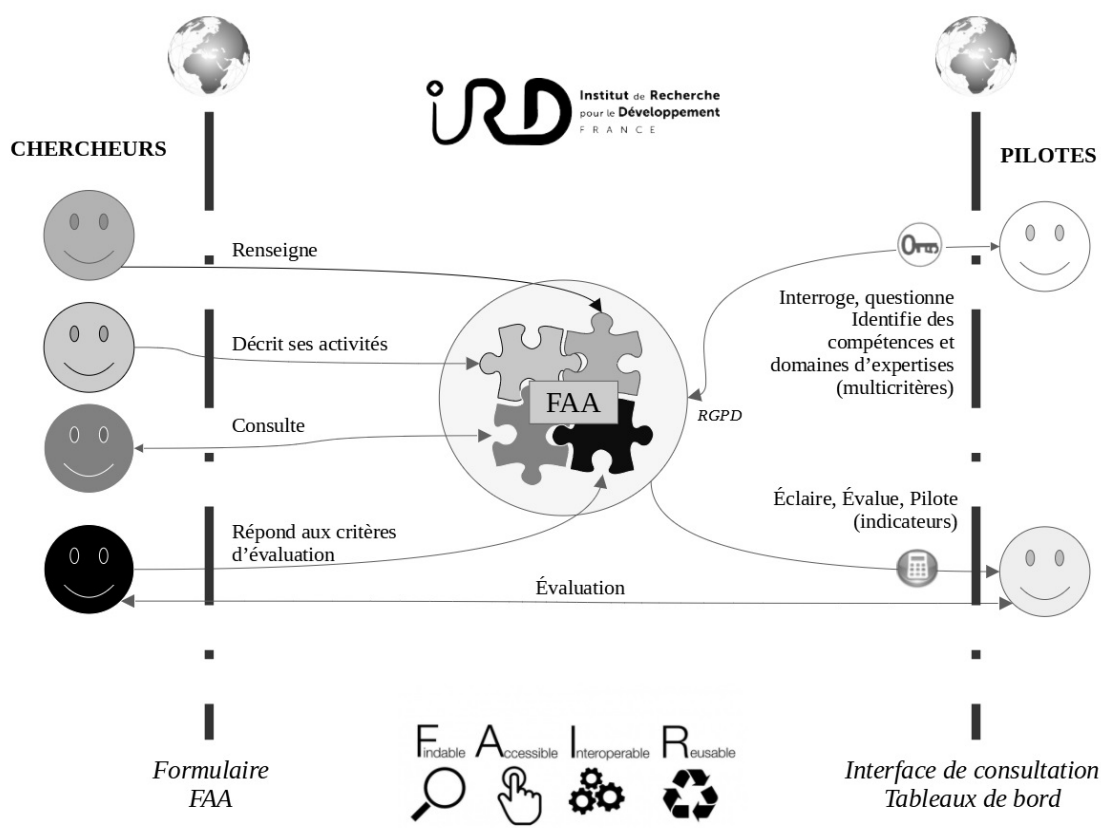

## 5.1 L'interface avec les chercheurs

### 5.1.1 *Le périmètre de confiance dans les données-sources et les indicateurs produits*

L'obligation, inscrite dans la réglementation relative à l'évaluation des chercheurs en EPST, est annuelle [10]. En revanche, l'évaluation par les pairs reste à ce jour biennale à l'IRD et le recueil des fiches annuelles d'activités de l'année N-1 sont déposées en ligne pendant la période d'évaluation biennale ou d'avancement l'année N, sous format Excel de 2017 à 2021. Les périodicités varient selon les établissements mais aucun ne programme d'évaluation individuelle des chercheurs à l'échelle annuelle comme cela est le cas pour l'évaluation des ingénieurs et des techniciens par leurs responsables hiérarchiques. Le caractère « obligatoire », mal connu, reste confus pour une partie des chercheurs au moment de réaliser l'exercice puisqu'ils sont sollicités une année sur deux pour le réaliser en-dehors du contexte de leur propre évaluation par les commissions (71 à 83% de la population de chercheurs et chercheuses titulaires ont décrit leurs activités annuelles 2016 à 2020 pendant les périodes d'évaluation 2017 à 2021). Le volume global de fiches recueillies reste néanmoins constant, autour de 600 personnes, sur un périmètre total allant de 750 à 800 personnes environ concernées réglementairement selon les années.

Chaque année, un effort de communication est fait auprès de tous les chercheurs sur la fiche et permet de maintenir un volume de données-source le plus constant possible en rappelant l'aspect contraignant mais réglementaire de l'exercice. Une aide précise est fournie pour chacun et chacune pour aider à la compréhension des questions posées ou sur l'usage fait des données transmises. Des améliorations sont possibles chaque année. Parmi les différents échanges, des chercheurs expriment eux-mêmes des propositions d'évolution des supports ; notamment la formulation des questions pour les rendre plus proches de leur réalité quotidienne et plus pertinentes pour l'institut ou pour leur évaluation individuelle. Ce suivi précis pendant plusieurs années (voir Fig. n°2) est une base d'identification des évolutions d'interface et des actions nécessaires pour faciliter l'exercice par 100% des personnes réglementairement interrogées. Ainsi, la couverture complète de ces données déclaratives d'activités permettrait d'atteindre un *périmètre de confiance* de 100% dans le contenu de la base données en amont à la production d'indicateurs et chiffres-clés, sans ajouter de circuit plus contraignant de validation à celui réglementaire.

**Figure n°2** : *Évolution des typologies d'interactions* avec les personnels IRD interrogés annuellement pour déclarer leurs activités via l'interface de la fiche annuelle d'activités. De 2012 à 2015, tous les personnels techniciens, ingénieurs et chercheurs fonctionnaires étaient interrogés par un formulaire en ligne. En 2017, la fiche annuelle d'activités a été réduite pour les seuls chercheurs, au format Excel jusqu'en 2021 pendant les périodes de campagnes d'évaluation. Le changement d'interface engendre chaque fois davantage de questionnements sur le périmètre (« suis-je concerné ? ») et implique un effort supplémentaire d'aide aux utilisateurs.

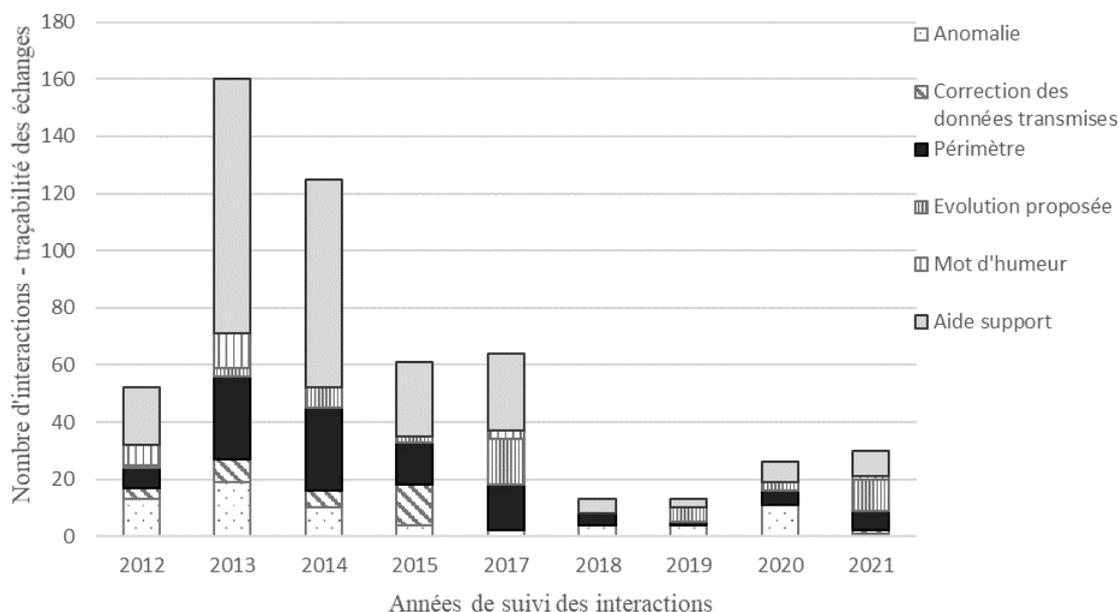

Source : IRD données de suivi Fiche annuelle d'activités, Sabine Tostain

### 5.1.2  Le périmètre des questions posées pour décrire les activités

L'ensemble des questions posées doivent permettre de répondre à des besoins d'identification de domaines et de sujets d'expertise, et à des besoins de calculs et de cartographies de volumétries pluriannuelles des différentes activités de chercheurs et chercheuses de l'institut. Pour obtenir des réponses exactes, les questions posées doivent :

- *évoluer avec les besoins* autant dans leurs périmètres (activités identifiées) que dans leurs formulations (textes et champs descriptifs des activités des chercheurs et chercheuses) : lesquelles abandonner ? Lesquelles ajouter ? Lesquelles modifier et ajuster ?

- *faciliter leur analyse et interprétation*, et donc limiter le « bruit » généré par des réponses apportées hors sujet (exemple : hors du périmètre attendu des questions posées) : quelles activités doivent être mieux délimitées ? Lesquelles génèrent des confusions au point de se confondre entre elles ? Lesquelles doivent être décrites spécifiquement, même si les réponses ne sont pas utilisées immédiatement ?

La fiche annuelle couvre aujourd'hui les activités suivantes, conformément aux missions des chercheur(euse)s d'un EPST [9], hors productions de type articles, ouvrages et chapitres d'ouvrages, littérature grise, identifiées par ailleurs dans une base documentaire interne à l'IRD :

- l'accompagnement de futures communautés scientifiques et l'implication dans des cursus universitaires ou auprès de professionnels ;
- le transfert, l'innovation et la diffusion de produits de la recherche, des savoirs et de la culture scientifique auprès des sphères scientifiques ou de la société civile (décideurs, entreprises, grand public, jeune public, médias, etc.) ;
- différentes formes d'expertises scientifiques ;
- le montage de projets et la recherche de financements.

Pour illustrer quelques changements marquants dans l'évolution des questions posées aux chercheurs, deux exemples : l'un sur une nouvelle question posée pour de nouveaux besoins d'observation et valorisation, le second pour réduire le bruit dans des réponses apportées et d'adapter la question au vocabulaire plus naturel des chercheurs.

#### 5.1.2.1  Exemple de nouvelle question posée

En 2011, aucune source n'existait en-dehors des outils de gestion contractuelle et financière pour cartographier l'effort de la communauté des chercheurs engagés en continu pour monter et construire des projets scientifiques avec des partenaires du Sud ; activité maintes fois mentionnée comme chronophage et peu valorisée dans l'évaluation. La « sphère » de bailleurs à laquelle s'adressaient les chercheurs était inconnue en-dehors des plus courants comme la Communauté européenne ou la Banque Mondiale par exemple. Aucune donnée d'observation n'existait donc non plus pour comprendre, identifier des leviers éventuels en terme d'appui nécessaire auprès des équipes et unités de recherche, et permettre d'augmenter – si besoin - leurs ressources propres et de rayonner davantage à l'international dans le cadre de la coopération scientifique. Une question a donc été formulée spécifiquement pour la première fois en 2021 sur le sujet des projets scientifiques, acceptés ou non.

Des données sur dix années sont aujourd'hui disponibles pour identifier les moyens déployés par les chercheurs de l'IRD et sur tous les sujets de recherche dans leurs environnements professionnels et communautés scientifiques respectifs. Des analyses pluriannuelles permettent notamment d'observer l'importance croissante de certains bailleurs étrangers, du Nord comme du Sud, et ce, selon les univers disciplinaires et les géographies d'intervention prévues dans lesdits projets. A l'échelle du chercheur, on constate également une récurrence dans les projets déposés en réponse à des appels d'offres : un projet non retenu une année est souvent représenté l'année suivante au même bailleur ou auprès d'autres partenaires scientifiques ou financiers.

Une nouvelle question formulée a donc permis 1) aux chercheurs de valoriser une activité pivot à la poursuite de leurs recherches et 2) de produire de nouveaux indicateurs et des analyses nouvelles sur des informations jusqu'alors inconnues. Le spectre d'observations s'est élargi.

### 5.1.2.2 Exemple de question ajoutée pour réduire le « bruit » dans les données-réponses

Améliorer la qualité des données-réponses aux questions formulées et réduire les difficultés d'interprétation peut se faire en identifiant les sources de confusion. Ces incompréhensions sont souvent générées par des concepts ou des vocabulaires utilisés communs mais recouvrant des périmètres différents d'activités selon l'interlocuteur. Identifier les zones de recouvrement est alors utile : poser deux questions au lieu d'une permet d'isoler le « bruit » généré par confusion de périmètres de l'unique question initiale. La « qualité » des données s'en trouve améliorée.

Depuis 2012, deux questions sont formulées dans la fiche annuelle d'activité pour décrire les activités en lien avec la diffusion d'une « culture scientifique » auprès du grand public ou des jeunes. Le service compétent à l'IRD accompagne et met en œuvre des moyens spécifiques sur ce qui est défini par de la « médiation » scientifique : de l'action et des rencontres avec le grand public ou le jeune public, et la construction et diffusion d'outils pédagogiques pour ces mêmes publics. En traitant annuellement les retours et les informations recueillies par la fiche annuelle, les chercheurs décrivaient pour leur part, également dans cette rubrique d'activités réalisées, toutes leurs interventions dans les médias ou pour des médias (télévision, radio, chaîne YouTube, etc.). Ces derniers touchant le grand public ou les enfants, il s'agissait donc pour eux de « rencontres » ou de « supports pédagogiques » pour la sphère non scientifique et donc de « vulgarisation » pour le grand public et la société civile. La présidence de l'institut souhaitant conserver ces éléments d'observation, en tant que source au rayonnement public de l'IRD et comme une activité à valoriser et identifier, une question a donc été ajoutée spécifiquement pour ces « rencontres avec les médias » à côté des actions grands publics et des outils pédagogiques afin de satisfaire tous les besoins. Un nouveau contenant est donc prévu depuis 2021 et réduit le bruit induit auparavant dans les autres contenus ; le tout en conservant la rubrique la plus englobante possible du concept inchangé de diffusion de la « culture scientifique ».

La précision des observations d'une activité tend donc à s'améliorer tout en conservant le périmètre familier des chercheurs et la correspondance de langage adéquate, et ce, malgré les différences connues avec le périmètre initial défini par le service-métier au siège de l'institut formulant seul les questions initiales.

## 5.2 L'interface avec les pilotes scientifiques

Afin de partager avec le plus grand nombre une partie des données présentes en base de données et permettre la construction autonome d'indicateurs agrégés selon des critères fréquents, une interface de consultation annuelle et pluriannuelle a été construite en interne à l'IRD (voir Fig. n°3). Les sujets choisis sont ceux faisant l'objet de questions fréquentes de la part des services centraux et de la gouvernance, et qui, également, sont les mieux maitrisés : les enseignements et formations continues dispensés, les étudiants et les doctorants encadrés ou dirigés, les actions grand public ou pour les jeunes et les outils pédagogiques conçus, les responsabilités pédagogiques dans des cursus d'école doctorale ou de master (ou équivalents à l'étranger).

L'objectif principal était de rendre visible la base de données existante et de répondre aux questionnements légitimes des chercheurs fournissant les données-sources : « à quoi cela sert de remplir la fiche annuelle d'activités ? ». Hormis l'aspect réglementaire pour les chercheurs directement concernés par l'exercice, cette mise en œuvre de tableaux de bord multicritères, dynamiques et accessibles par simple navigateur et identification, avait pour second impact attendu d'éclairer davantage de pilotes de la Science au sein de l'Institut (Présidence, départements scientifiques, directions d'unités, représentations, directions centrales, chefs de projets, etc.). Une centaine d'utilisateurs ont aujourd'hui accès à ces indicateurs (voir Fig. n°3) et sont autonomes pour interroger les données-sources non nominatives sur ces activités décrites. Différentes demandes d'amélioration ont été recensées, prouvant l'intérêt pour ce type d'interface et appétence sur ces données par les utilisateurs.

Parmi les évolutions identifiées, sont programmées :

- l'ajout d'indicateurs d'activités supplémentaires déjà décrites dans la fiche annuelle (expertises, montages de projets scientifiques, savoirs et instruments innovants) ;
- l'ajout de nouveaux critères de filtres disponibles (par commissions d'évaluation, par enjeux scientifiques ou disciplines de recherche, etc.).

Ces tableaux de bord sont par ailleurs eux-mêmes sources d'autres indicateurs ou d'autres interfaces de présentation et consultation des données, conçues depuis ou en construction pour la gouvernance, pour les directions d'unités et pour les représentations de l'IRD hors métropole.

Les cahiers de l'école qualité et responsabilité sociétale en recherche et en enseignement supérieur – Septembre 2021 – *in press 2023*

**Figure n°3** : *Illustration des tableaux de bord dynamiques* construits pour visualiser plusieurs indicateurs pluriannuels d'activités des chercheur(euse)s de l'IRD de 2016 à 2020 sur une partie des sujets communément consultés. Les données résultent du traitement et consolidation des déclarations par la fiche annuelle d'activités. L'interface de restitution est accessible via intranet après attribution de droits d'accès (une centaine d'utilisateurs à date) ; elle a été construite en interne et fera l'objet d'évolutions et enrichissements pour s'adapter aux besoins

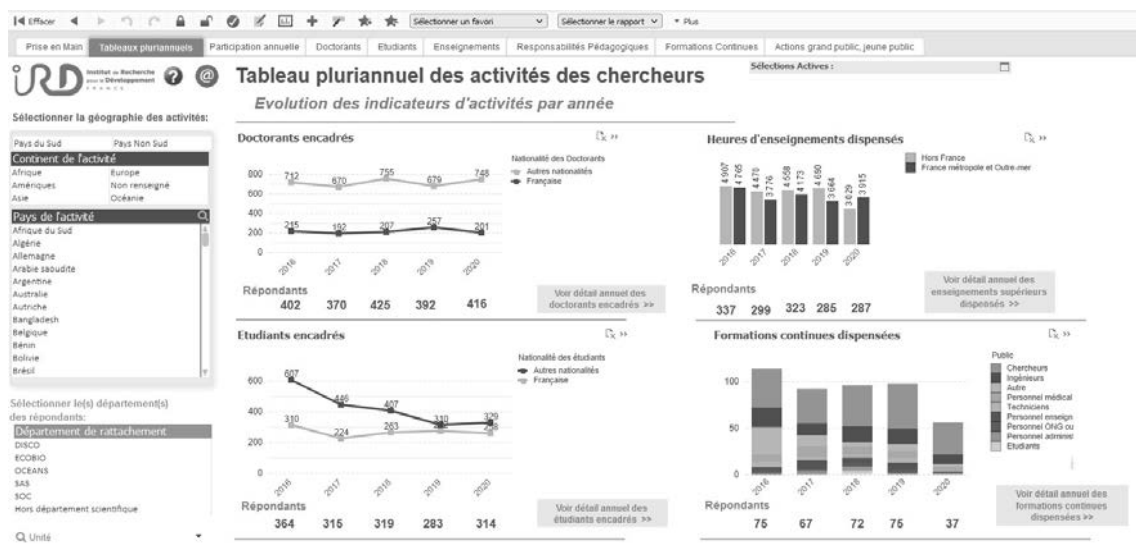

internes de pilotage et à la simplification des flux de données mises à jour annuellement.

## 6 Les améliorations continues : retours d'expérience du projet FAA

Les principaux retours d'expériences du projet en terme d'amélioration continue sont :

- l'identification des besoins de changements pour permettre de saisir les opportunités dès qu'elles se présentent ;
- l'analyse transversale et stratégique en amont à la planification adaptative des actions à mener pour piloter le projet et accompagner les changements en conséquence sur la base des risques maitrisables.

### 6.1 Les clés de changements : des opportunités à saisir

Etant donné le périmètre et la construction somme toute assez jeune de ce « système » décrit à l'IRD autour de la fiche annuelle d'activités, un changement de personnes-ressources dans des services impliqués dans le projet, des changements structurels d'organisation des services ou de la gouvernance, les évolutions technologiques ou l'apparition de nouveaux enjeux d'observation institutionnels sont de potentiels freins aux changements. L'observation et l'analyse de toute opportunité interne d'amélioration - pour tout ou partie du système - s'avère donc nécessaire pour progresser étape par étape en posant des « jalons » solides pour consolider le système, d'ampleur croissante pour le cas de celui couvert par le projet FAA à l'IRD (voir Fig. n°4).

**Figure n°4** : Schéma d'amélioration continue dans le temps pour la production et la maîtrise des données d'activités des chercheurs de l'IRD selon la norme ISO9001 et la méthode PCDA (source IRD Sabine Tostain).

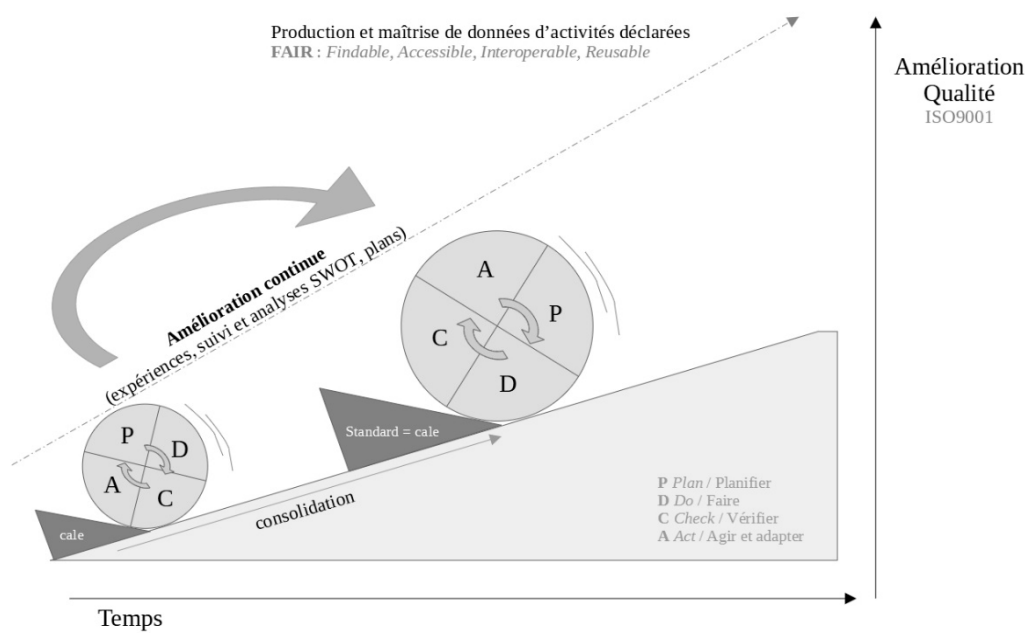

En 2016 par exemple, après une année blanche de recueil de données à la suite d'un changement d'équipe de gouvernance et d'abandon d'un précédent formulaire en ligne, le recueil d'informations des fiches annuelles d'activités des chercheurs a repris sous un format Excel. Stabilisant années après années la formulation des questions posées aux chercheurs, la base de données pluriannuelle résultante, très riche, restait en 2018 mal connue des services en interne et sous-exploitée en terme de production d'indicateurs. En conséquence, les chercheurs eux-mêmes restaient sceptiques quant à l'intérêt réel de réaliser l'exercice compte tenu d'un usage faible ou méconnu de leurs déclarations, et ce malgré l'aspect réglementaire annuel. Parallèlement à l'IRD, de nouveaux outils numériques et référentiels internes tentaient d'être mis en place. Fin 2018, avec peu de ressources, a été identifiés conjointement l'opportunité de construire des tableaux de bord ergonomiques sur des données existantes et certaines activités décrites ont été utilisées pour constituer une interface-pilote comme moyen d'aide au pilotage. Cette opportunité numérique permettait de répondre à deux principaux freins identifiés : la question récurrente de « à quoi cela sert de remplir des tableaux pour détailler ces activités chaque année ? », et de donner de la visibilité aux données et indicateurs produits tout en rendant « appétences » les données existantes répondant à plusieurs besoins exprimés. Cette interface a été le second jalon solide posé après la stabilisation du périmètre des questions.

Compte tenu des besoins de transversalité des données-sources, des besoins de réduction des temps de traitements des données brutes pour être exploitables et partagées, et pour augmenter le périmètre de confiance et le nombre de répondants à cette obligation réglementaire, le projet FAA a été recentré ensuite fin 2020 sur l'interface avec les chercheurs et chercheuses : *l'ergonomie et l'interface de la « fiche annuelle d'activités » avec les chercheurs*.

L'opportunité de la construction de la feuille de route *Science ouverte* à l'IRD, concomitante au déblocage de moyens financiers pour *moderniser* et *simplifier* les interfaces scientifiques dans le cadre du Schéma directeur du numérique de l'IRD ont été des facteurs déclenchants de l'amélioration de l'interface du formulaire en lieu et place du support Excel pour la fiche annuelle. D'un côté des indicateurs et tableaux de bord faciles d'accès et partagés étaient préexistants et mis à jour chaque année pour illustrer certains usages possibles pour le pilotage scientifique. De l'autre côté, la saisie d'informations-sources et mise en qualité des données recueillies nécessitait d'être simplifiée et mieux maîtrisée (voir les cartouches des *forces* et des *opportunités* dans la Fig. n°5).

Les retours d'expériences antérieures et le suivi annuel des interactions avec les chercheurs (voir Fig. n°2) ont nourri la liste initiale des besoins identifiés et décrits dans le cahier des charges des développements. Plus de 25 ateliers ont été menés début 2021 en interne avec chaque partie prenante et collégialement (les « directions-métiers »). Un *groupe pilote* d'une vingtaine de chercheurs et chercheuses volontaires – toutes disciplines scientifiques représentées - a également été constitué et mobilisé dès la définition des besoins fonctionnels du formulaire annuel d'activités pour s'assurer d'intégrer également leurs besoins et leurs remarques.

## 6.2   Les clés de la planification : l'analyse initiale par SWOT[4]

Le « système » global de la fiche annuelle d'activités est en interactions fortes avec d'autres périmètres fonctionnels ou techniques en interne à l'IRD (par exemples la définition des enjeux institutionnels ou la construction et le suivi du Contrat d'objectifs, le processus de l'évaluation biennale, ou celui de l'urbanisation du système d'informations de l'IRD, le suivi de contrats de recherche, l'identification et le suivi des produits ou des innovations de la recherche, etc.).

---

[4] Forces, Faiblesses, Opportunités et Menaces (en anglais : *Strengths, Weaknesses, Opportunities, and Threats* – SWOT)

Les cahiers de l'école qualité et responsabilité sociétale en recherche et en enseignement supérieur – Septembre 2021 – *in press 2023*

**Figure n°5** : Analyse SWOT résumée du projet FAA sur le volet d'évolution de l'interface avec les chercheurs et des contenus de la fiche annuelle d'activités à l'IRD - phase de cadrage (source IRD Sabine Tostain, version de mars 2021).

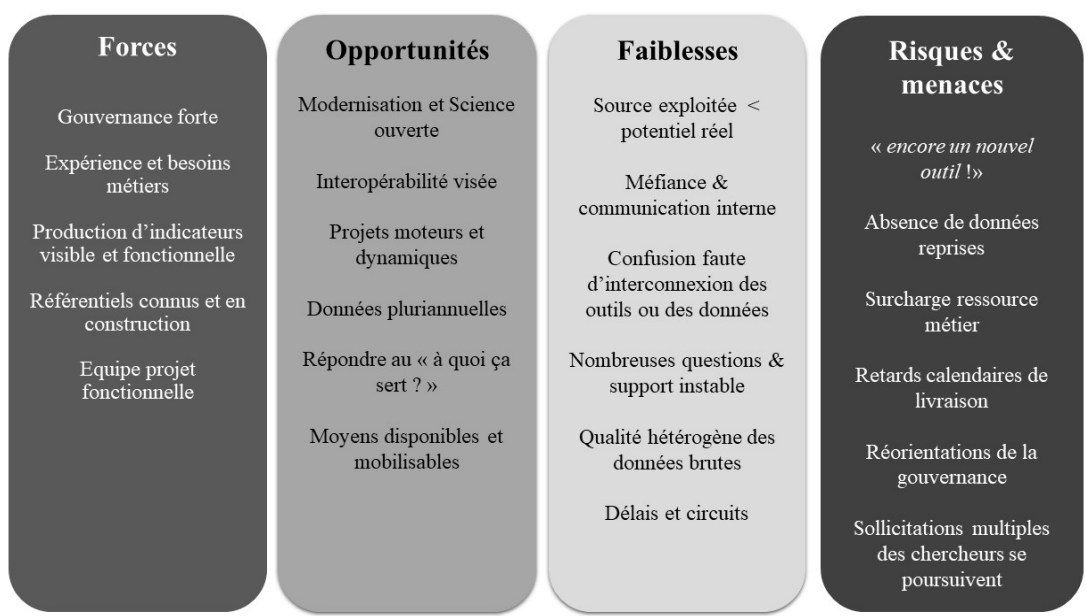

Les plans d'actions et la programmation annuelle ne sont pas repris en détails dans le présent article mais des extraits de contenus sont présents ci-après à titre d'illustration corrélée au SWOT initial :

- le niveau des risques et menaces du projet selon les impacts identifiés (Tableau n°1), avec pour chaque, des actions spécifiques pour le gérer et réagir dans un *plan de gestion des risques et ROI*[5]. Le suivi de l'apparition croissante ou critique d'un risque implique aussi de reprogrammer ou d'adapter des actions, notamment dans celles du plan de communication et d'accompagnement au changement,

- les différents axes et ambitions associées aux différentes actions du *plan de communication et d'accompagnement au changement*, construit sur 4 objectifs-cibles de communication (anticiper et prévenir, informer et accompagner, diffuser et promouvoir, valoriser) (Tableau n°2).

---

[5] ROI : retour sur investissement

Les cahiers de l'école qualité et responsabilité sociétale en recherche et en enseignement supérieur – Septembre 2021 – *in press 2023*

**Tableau n°1** : *extrait synthétique des risques identifiés* (R) du projet FAA au 1er septembre 2021, classés par niveaux résultats à une matrice probabilité (P) x gravité (G). Le suivi des risques permet d'agir ou réagir en fonction de leur apparition ou de leurs évolutions, en adaptant le plan d'actions associées pour les maîtriser (source IRD Sabine Tostain).

| RISQUES & MENACES IDENTIFIES (R) | IMPACTS | NIVEAU DE RISQUE (P*G) / *1 très élevé* | | |
|---|---|---|---|---|
| | | 1 | 2 | 3 |
| Retards calendaires de mise en production et lancement du formulaire FAA au-delà du 1er trim. 2022 | Durée trop courte pour renseigner 2021 avant recueil suivant 2022 Accès aux archives FAA non fonctionnel pour servir aux utilisateurs dans leurs autres démarches, notamment d'évaluation ou candidature, etc. Ajout de complexité (2 circuits et supports en parallèle à remplir) pour l'utilisateur | X | | |
| | Augmentation du mécontentement et mots d'humeurs Mauvaise image de la nouveauté et perte d'intérêt | X | | |
| Données antérieures non reprises ou prêtes à l'ouverture | Promesses fonctionnelles non tenues et perte de sens du projet | | X | |
| Surcharge ressources humaines de l'équipe-projet ou copilotes des données | Données non traitées à la MEPR ou non orientées vers autres parties prenantes métier identifiées | X | | |
| | Données non traitées ou maitrisées en base (circuits, contenus, qualité) | X | | |
| « encore un nouvel outil ! » - mauvaise anticipation de l'accompagnement au changement | Baisse du périmètre de confiance de la représentativité des données et indicateurs produits | | X | |
| | Justifier au lieu de communiquer et informer | | X | |
| | Baisse du nombre de contributeurs annuel Mauvaise appropriation par les chercheurs Ne pas respecter les engagements de partage et ouverture des données Ne pas simplifier les processus internes | | | X |
| Sollicitations multiples des chercheurs se poursuivent | Communication dégradée ou absente entre pilotage et utilisateurs ; rupture de confiance et incompréhension des utilisateurs sur les flux de données ; transparence et transversalité annoncées & engagements IRD / SDN non respectés | | X | |
| Changements de stratégie concernant FAA | non maîtrisable | | X | X |

Les cahiers de l'école qualité et responsabilité sociétale en recherche et en enseignement supérieur – Septembre 2021 – *in press 2023*

**Tableau n°2** : *plan d'actions de communication synthétique* pour le projet FAA à l'IRD sur quatre objectifs principaux : valoriser, promouvoir, faire savoir et accompagner. Plusieurs actions sont identifiées selon les thématiques de communication (ou sujets). Une partie des actions ou de ces thématiques de communication répond parfois à plusieurs objectifs (source et conception IRD Sabine Tostain).

| Objectifs | Thématiques de communication | Nombre d'actions identifiées |
|---|---|---|
| **Valoriser** (montrer l'importance des activités riches et internationales des chercheurs IRD + répondre à la question "à quoi ça sert ?") | Afficher publiquement les résultats ou certaines données | 2 |
| | Eclairer les pilotes | 1 |
| | Encourager les chercheurs à contribuer | 1 |
| | Partager des données sécurisées | 4 |
| | Valoriser les données | 3 |
| | Valoriser l'effort de saisie des chercheurs via le formulaire et la centralisation des données | 1 |
| | Nombre d'actions pour « Valoriser » | 12 |
| **Promouvoir** (illustrer les évolutions et innovations internes en terme d'outils modernisés et de pilotage éclairé) | Afficher les nouveautés | 1 |
| | Afficher les *success stories* | 1 |
| | Souligner les innovations | 1 |
| | Nombre d'actions pour « Promouvoir » | 3 |
| **Faire savoir** (informer synthétiquement sur les activités du siège en lien avec les activités réelles des chercheurs) | Encourager les chercheurs à contribuer | 1 |
| | Informer sur la production d'indicateurs | 1 |
| | Informer sur les changements | 1 |
| | Informer sur les processus et pilotage des données | 1 |
| | Rendre accessible les tableaux de bord existants | 1 |
| | Sensibiliser la communauté IRD | 2 |
| | Nombre d'actions pour « Faire savoir » | 7 |
| **Accompagner** (accompagner aux changements et aider les utilisateurs et consommateurs de données, rappeler droits et devoirs de chacun) | Aide aux utilisateurs | 7 |
| | Anticiper les changements | 1 |
| | Encourager les chercheurs à contribuer | 1 |
| | Informer sur les changements | 1 |
| | Respecter la loi et protéger les données | 1 |
| | Rendre accessible le nouvel outil | 1 |
| | Nombre d'actions pour « Accompagner » | 12 |

# 7 Conclusion

Au-delà des flux de données-métiers et de la conception d'outils, les enjeux de la transformation pour améliorer du la production d'indicateurs d'activités des chercheurs nécessitent un véritable accompagnement aux changements : un dialogue interservices et une culture du « partage », avec une « culture qualité » de la donnée et une « culture d'indicateurs de pilotage » par tableaux de bord.

Les retours d'expériences au fil des années et le pilotage continu du projet au niveau de la direction scientifique de l'IRD ont permis de faire évoluer l'ensemble du système pour les deux typologies d'utilisateurs ou « clients » pour répondre à leurs besoins respectifs. Ces évolutions sont réalisées de manière asynchrone mais contribuent, dans le temps, à consolider et fiabiliser tout ou partie du système. Parmi les points forts notables en conclusion de ce partage d'expériences : 1) considérer le chercheur lui-même comme client du système et non comme seul fournisseur d'informations, 2) intégrer des chercheurs dans le pilotage du projet et l'identification des axes de changements nécessaires, 3) une ou des personnes ressources mobilisée en continu sur plusieurs années au pilotage du projet et/ou le management qualité avec une vue transversale à tout le système, et enfin 4) mobiliser et animer le réseau des directions-métiers des services centraux, parties prenantes les plus actives, comme *acteurs* dans le pilotage des données et des interfaces, en plus d'être par ailleurs des *consommateurs actifs* des indicateurs produits sur cette base de données interne constituée. Ce dernier point fort permet d'initier, en interne, la formation d'un groupe de « copilotes » éclairés et sensibles à la qualité des informations et à leurs différents usages.

La majorité des indicateurs les plus communs pour décrire « *les activités de la recherche* » sont construits sur des données considérées comme « sures » ou « vraies » car mesurables aujourd'hui par les services centraux des établissements. Cette « assurance-qualité » est appuyée par l'usage de flux institutionnels dans le traitement des informations, après une ou plusieurs étapes de validation, souvent sur des sources documentaires (devis, factures, diplômes, attestations, décisions signées, etc.). Les indicateurs les plus courants et disponibles sont par conséquent des indicateurs relevant de la gestion de ressources, humaines ou financières. Ils résument une certaine forme de performance entre dépenses, investissements et les résultats obtenus. Donner de la valeur à de l'information déclarative et produire des indicateurs robustes d'activités des chercheurs et chercheuses, pour une aide au pilotage scientifique complémentaire à celle fournie par les indicateurs de gestion préexistants : là résident les enjeux de l'amélioration continue du système actuellement construit et intégré à partir de la « fiche annuelle d'activités (FAA) » à l'IRD.

Les cahiers de l'école qualité et responsabilité sociétale en recherche et en enseignement supérieur – Septembre 2021 – *in press 2023*

# 8   Bibliographie et remerciements

Sabine TOSTAIN, cheffe de projet et référente Qualité à la Mission d'évaluation et programmation de la recherche
Institut de Recherche pour le Développement (IRD)
Bât. Le Sextant - 44 Boulevard de Dunkerque - CS 90009 - 13572 Marseille cedex 02
Marseille, France
sabine.tostain@ird.fr